\documentclass[preprint,aps,amsmath,eqsecnum]{revtex4}
\usepackage{bm}
\usepackage[dvips]{graphicx}

\begin{document}

\title{Exciton correlations in coupled quantum wells and their luminescence blue shift}

\author{B. Laikhtman and R. Rapaport}
\affiliation{Racah Institute of Physics, Hebrew University, Jerusalem 91904
Israel}

\begin{abstract}
In this paper we present a study of an exciton system where electrons and holes are confined in double quantum well structures. The dominating interaction between excitons in such systems is a dipole - dipole repulsion. We show that the tail of this interaction leads to a strong correlation between excitons and substantially affects the behavior of the system. Making use of qualitative arguments and estimates we develop a picture of the exciton - exciton correlations in the whole region of temperature and concentration where excitons exist. It appears that at low concentration degeneracy of the excitons is accompanied with strong multi-particle correlation so that the system cannot be considered as a gas. At high concentration the repulsion suppresses the quantum degeneracy down to temperatures that could be much lower than in a Bose gas with contact interaction. We calculate the blue shift of the exciton luminescence line which is a sensitive tool to observe the exciton - exciton correlations.
\end{abstract}

\maketitle

\section{Introduction}

A very active investigation of excitons in coupled quantum wells for more than
two decades was first motivated by the possibility to reach Bose condensation
and superfluidity in this system. Further experiments discovered a very large
number of related phenomena and quite rich physics of the system (see
Refs.\cite{Snoke02S,Butov04,Stern08} and references therein). Theory predicts
the existence of many phases with different and unusual
properties.\cite{Lozovik96,Fernandez-Rossier96,Ben-Tabou03,Astrakharchik07} The
most expected and searched for is the settling in of coherence in such 2D
exciton systems. A coherence of the exciton Bose condensate has to reveal
itself in some coherent properties of the exciton luminescence. Investigation
of the luminescence led to discovery of not only its coherence
\cite{Richard05,Yang06} but also a number of patterns not completely understood
so far \cite{Butov02,Snoke02N,Rapaport04,Gorbunov06}.

\begin{figure}
\includegraphics[scale=0.5]{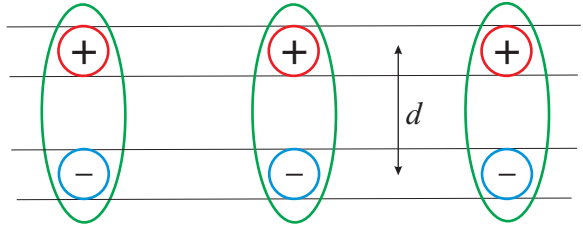}
\caption{\label{fig:X_in_CQW}Excitons in coupled quantum wells. $d$ is the
separation between the centers of the wells (given $d$ the dependence of the
exciton - exciton interaction on the well widths is very weak
\cite{Ben-Tabou01}). $d$ and the average separation between excitons are
assumed to be larger than the exciton radius $a_{X}$.}
\end{figure}

A substantial role in these phenomena is played by the interaction between
excitons. Typically, interaction between bosons ($^{4}$He atoms and
alkali-atoms) is of a short range and the theory of non-ideal Bose gas has been
developed for contact interaction. \cite{Huang,lifshits_sp} In coupled quantum
wells where the electrons and holes are separated in the two adjacent layers,
all the indirect, dipolar excitons that are formed by the coulomb binding of
pairs of these spatially separated electrons and holes are polarized in the
same way and their interaction is mainly dipole -- dipole repulsion,
Fig.\ref{fig:X_in_CQW}. Contrary to the contact interaction the dipole --
dipole interaction has a significant tail and due to this tail the exciton gas
in some respects is dramatically different from Bose gas with contact-like
interactions.

The dipole -- dipole repulsion increases the exciton energy and leads to a blue
shift of its luminescence line. The blue shift is usually evaluated as an
average value of the exciton interaction energy with other excitons and can be
easily obtained with the help of the mean field approximation that produces
"plate capacitor formula"\cite{Butov99}
\begin{equation}
E_{int} = \frac{4\pi ne^{2}d}{\kappa} \ ,
\label{eq:1}
\end{equation}
where $n$ is the exciton 2D concentration, $d$ is the separation between the
centers of the wells, and $\kappa$ is the dielectric constant. This formula can
be understood in the following way. Equal concentration $n$ of electrons and
holes in two wells creates potential difference $\Delta\phi=4\pi ned/\kappa$
between them. Creation of one more indirect exciton in these wells requires
transfer of an electron or a hole from one well to the other. In the presence
of other excitons this increases the necessary energy by $e\Delta\phi$ that
gives Eq.(\ref{eq:1}). This expression is typically used in experiments for an
estimate of the exciton concentration from the measured blue shift of the
luminescence.

Recently Zimmermann and Schindler \cite{Zimmermann07} noticed, however, that
dipole -- dipole repulsion leads to a significant exciton pair correlation. The
repulsion prevents excitons to come very close to each other and creates a
depleted region around each exciton. The pair correlation leads to a reduction
of the coefficient in Eq.(\ref{eq:1}) by about 10 times (depending on the
excitons temperature).\cite{Zimmermann07} This means that previous experimental
estimates of the exciton concentration based on the capacitor formula
underestimated the concentration by up to 10 times!

It makes sense to note that the reduction of the pair correlation function to
zero at small distance in a Bose gas with repulsion has been noticed long ago
and used in the construction of a many particle variational wave
function.\cite{Dingle49,Jastrow55} It is well known in exact solutions for one
dimensional Bose gas.\cite{Lieb63,Yang69,Sutherland71,Sykes08} In the exciton
gas with dipole -- dipole repulsion this behavior of the pair correlation
function was noticed by Astrakharchik et al.\cite{Astrakharchik07}. This
behavior was also used by Kash et al to explain a narrowing of the exciton
luminescence line at low temperature.\cite{Kash91}

The significant pair correlation between excitons substantially affects our
understanding of the behavior of this system. This motivated us to develop a
general picture of the interaction and correlation between excitons in the
whole temperature -- concentration plane. This is the subject of this paper. We
show that Eq.(\ref{eq:1}) has a very limited region of applicability and
limited accuracy. With a decrease of the temperature and an increase of the
concentration the role of the pair correlation becomes crucial for
interpretation of such phenomena as the blue shift. It appears also that the
degeneracy of the exciton system at any small concentration is accompanied by
setting in of not pair but multi-particle correlation characteristic for
liquids. In other words, a consistent theory of a quantum coherent state has to
include multi-exciton correlation at any dilute gas. More than this, at higher
concentrations the dipole -- dipole repulsion reduces the overlap of the wave
functions of different excitons. As a result the exciton system can be
considered as classical liquid down to temperature well below than the
temperature of quantum degeneracy of a system with contact interaction.

Our main purpose is to develop a qualitative understanding of the structure of
the system of interacting bosons. Therefore we don't pursue a high accuracy of
the results but instead use simplified models and qualitative arguments.
Although the results of such an approach are really accurate only in extreme
cases with respect to some large or small parameters, they allow us to make
analytical calculations and produce a very clear physical picture of relevant
phenomena. Availability of such a picture facilitates precise calculations when
they are necessary.

In the next section we consider in detail the applicability of the mean field
approximation that neglects any correlation between excitons. An exciton gas at
lower temperatures where strong pair correlation is important but quantum
correlations are still negligible is considered in Sec.\ref{sec:cxg}. Further
reduction of temperature at low exciton concentrations, when the exciton
wavelength becomes larger than the characteristic scale of exciton - exciton
interaction, leads to an important role of quantization in the  exciton -
exciton scattering while the exciton gas itself is still statistically
non-degenerate. This situation is considered in Sec.\ref{sec:xgqs}. In
Sec.\ref{sec:lsxs} we consider the situation where multi-exciton correlation is
important. This happens at low enough temperature in a dilute system and in a
more wide range of temperatures in a dense system. It appears that with a
temperature decrease in a dilute system degeneracy is accompanied by a set in
of a multi-exciton correlation. Contrary, in a dense system the classical
multi-exciton correlations appear to be more important for the blue shift than
the quantization of exciton dynamics. In the last section we discuss the
obtained results.

The discussion of exciton correlation in different parts of the concentration -
temperature plane involves quite a large number of physical parameters. To
facilitate reading of the paper we present here the list of main parameters
with their physical definitions. Exact mathematical definitions, if necessary,
will be given as soon as the corresponding parameter comes into the discussion.

\begin{table}
\caption{\label{tab:parameters}The main parameters characterizing an exciton gas in
coupled quantum wells} \centering
\begin{tabular}{||c|l||c|l||}
\hline
 $d$                & Separation between electron and hole wells
    & $a_{B}$       & Bohr radius with a reduced electron - hole mass     \\
 $m_{e}$, $m_{h}$   & Electron and hole masses
    & $b$           & Bohr radius with an exciton mass                     \\
 $m$                & Reduced electron - hole mass
    & $a_{X}$       & Exciton radius                                    \\
 $M$                & Exciton mass
    & $r_{0}$       & Average minimal distance between excitons         \\
 $\epsilon_{b}$     & Exciton binding energy
    & $k_{T}$       & Thermal wave vector of an exciton                    \\
 $n$                & Exciton concentration
    & $U(r)$        & Interaction energy between excitons               \\
 $E_{int}$          & Average interaction energy of a single exciton
    & $g(r)$        & Exciton pair correlation function                 \\
\hline \hline
\end{tabular}
\end{table}

\section{Mean field approximation}
\label{sec:mfa}

\begin{figure}
\includegraphics[scale=0.5]{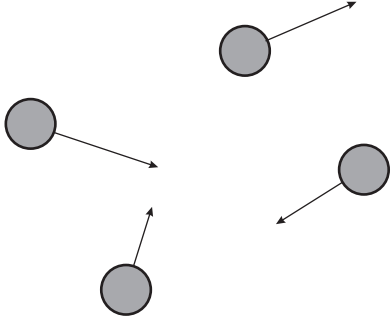}
\caption{\label{fig:Random}Uncorrelated motion of excitons. The radius of each
circle depicts the characteristic scale of the repulsive potential $d$.
A random motion takes place only at low concentrations, Eq.(\ref{eq:mfa.3}), and
rather high temperatures, Eq.(\ref{eq:mfa.5}).}
\end{figure}

The interaction energy between excitons in coupled quantum wells is
\begin{equation}
U(r) = \frac{2e^{2}}{\kappa}
    \left(\frac{1}{r} - \frac{1}{\sqrt{r^{2} + d^{2}}}\right)
\label{eq:mfa.1}
\end{equation}
According to Refs.\cite{Lozovik96,Zimmermann07} the attractive Van der Waals
interaction and the exchange interaction are small in practically important
values of $d$. The simplest way to obtain the average interaction energy is to
assume that all excitons are distributed randomly and independently of each
other with an average concentration $n$, see Fig.\ref{fig:Random}. Then the
average number of excitons in an area element $d^{2}\bm{r}$ is $nd^{2}\bm{r}$
and the average interaction energy is
\begin{equation}
E_{int} = \int U(r) n d^{2}r = \frac{4\pi ne^{2}d}{\kappa} \ .
\label{eq:mfa.2}
\end{equation}
This result means that the main contribution to $E_{int}$ comes from the
interaction between excitons at distance of the order of $d$: $(e^{2}/\kappa
d)(n\pi d^{2})=\pi ne^{2}d/\kappa$.

Eq.(\ref{eq:mfa.2}) is valid only under a few assumptions. The first is that
the average distance between excitons has to be larger than $d$. In other
words, the concentration cannot be too large,
\begin{equation}
nd^{2} \ll 1 \ .
\label{eq:mfa.3}
\end{equation}
In the opposite case the electron -- electron and hole -- hole repulsion is
stronger than the electron -- hole attraction and it is hardly possible to
expect a stable exciton phase of the
system.\cite{Littlewood96,Palo02,Balatsky04,Joglekar06} In all following
considerations we assume that inequality Eq.(\ref{eq:mfa.3}) is met.
Practically, this limitation in coupled well structures with $d\gtrsim10$ nm
means than the concentration has to be much smaller than $10^{12}$ cm$^{-2}$.

The other main assumption used in the derivation of Eq.(\ref{eq:mfa.2}) is the
absence of any correlation between excitons. In reality, if the kinetic energy of the
relative motion of any two excitons is $E$ they can approach each other only to a
distance larger than $r_{0}(E)$ where $r_{0}$ is the root of the equation
\begin{equation}
U(r_{0}) = E \ .
\label{eq:mfa.4}
\end{equation}
The correlation between excitons can be neglected only if the region where the
correlation is important is very small, i.e., $r_{0}\ll d$. If the temperature
of the exciton gas is $T$ then $E\sim T$ and this condition can be written as
\begin{equation}
T \gg e^{2}/\kappa d \ .
\label{eq:mfa.5}
\end{equation}

Condition (\ref{eq:mfa.5}) is equivalent to $k_{T}\gg1/\sqrt{bd}$ where
$k_{T}=\sqrt{2MT}/\hbar$ is the exciton thermal wave vector and
$b=\hbar^{2}\kappa/Me^{2}$. The expression for $b$ differs from the Bohr radius
$a_{B}=\hbar^{2}\kappa/me^{2}$ only by replacement of the electron - hole
reduced mass $m=m_{e}m_{h}/(m_{e}+m_{h})$ with the exciton mass
$M=m_{e}+m_{h}$. In GaAs/AlGaAs structures the electron effective mass
$m_{e}=0.067$ and the hole effective mass at the bottom of hh1 subband in a
quantum well $m_{h}=0.14$. This gives $m\approx0.045$, $M\approx0.21$, and
therefore $a_{B}\approx14$ nm, and $b\approx3$ nm which allows us to assume in
further calculations that $b\ll d$. Therefore the condition (\ref{eq:mfa.5})
means also that $k_{T}\gg1/d$, i.e., the exciton wavelength is much smaller
than the characteristic length scale of the potential. This justifies a
classical consideration of the interaction between excitons. Also, this
inequality in combination with Eq.(\ref{eq:mfa.3}) leads to the inequality
$k_{T}^{2}\gg n$ which means that the exciton thermal wavelength is much
smaller than the average inter-particle distance and therefore the exciton gas
is non-degenerate.

It makes sense to note that a quantum mean field approximation also leads to
Eq.(\ref{eq:mfa.2}).\cite{Yoshioka90,Zhu95,Ben-Tabou01,Ivanov02} However, if
the exciton radius $a_{X}$ is of the order of or larger than $d$ the exchange
interaction also appears to be important.\cite{Ben-Tabou01} Variational
calculations for GaAs/AlGaAs structures with an exciton wave function
$\psi(r)=Ae^{-\sqrt{d^{2}+r^{2}}/2a_{X}}$ yields, for layers separation of
$d=10$ nm, 12 nm, and 14 nm the corresponding exciton radii of $a_{X}=8.7$ nm,
9.3 nm, 9.7 nm, and exciton binding energies of 4.7 meV, 4.2 meV, and 3.8 meV
respectively. That is although $d>a_{X}$ this inequality is not very strong and
there are quantum corrections to the interaction in
(\ref{eq:mfa.1}).\cite{Zimmermann07,Ben-Tabou01}

The exciton binding energy also puts an upper limit to the temperature where
the above mean field description is viable because at temperatures of the order
of or larger than the binding energy the majority of the excitons dissociate.
According to Eq.(\ref{eq:mfa.5}) the low limit for the temperature for $d=10$
nm, 12 nm and 14 nm is 54 K (4.7 meV), 48 K (3.8 meV) and 44 K (8.6 meV)
respectively.

The bottom line of these estimates is that Eq.(\ref{eq:mfa.2}) is a rather poor
estimate: it is really valid only in a the temperature range where a
significant part of excitons is dissociated.

To conclude this section we comment on a usage of the mean field approximation
in explaining phenomena other than the exciton luminescence blue shift. The
interaction of one of the particles with all others can be described with the
field created at the particle by the environment. In general, this field
fluctuates in time and from particle to particle due to different dynamics of
particles creating it. If the particle interacts simultaneously with many
others and they are not correlated then these fluctuations are cancelled and
their resulting amplitude is much smaller than the average value of the field.
This is the foundation that makes the mean field approximation valid. In the
exciton gas with dipole -- dipole interaction the situation is quite different.
The average interaction energy Eq.(\ref{eq:mfa.2}) is much larger than the
interaction between excitons at average distance between them: $U(n^{-1/2})\sim
e^{2}d^{2}n^{3/2}/\kappa$ and $U(n^{-1/2})/E_{int}\sim(nd^{2})^{1/2}\ll1$. This
means that the main contribution to $E_{int}$ comes from rear pairs of excitons
with the distance much smaller than the average one. The large amplitude of the
field fluctuations is also confirmed by the calculation of the average of the
interaction energy squared:
\begin{equation}
\overline{U^{2}} = \int U^{2}(r)n d^{2}\bm{r} =
2\pi\left(\frac{2e^{2}n^{1/2}}{\kappa}\right)^{2} \ln\frac{d}{r_{0}} \ .
\label{eq:mfa.6}
\end{equation}
Discarding the logarithm that comes from the cutoff of the minimal distance
between excitons we see that $E_{int}^{2}/\overline{U^{2}}\sim nd^{2}\ll1$.
(Under this condition $\overline{U^{2}}$ characterizes the luminescence line
width, but a discussion of this point is beyond the scope of the paper.)

The message is that while the mean field approximation for the average
interaction gives a correct result in the mentioned range of parameters, the
calculation of other quantities in this approximation can lead to large errors.
It is also important to keep in mind that in spite of the purely classical
arguments this statement is true also in the quantum limit.

\section{Classical exciton gas}
\label{sec:cxg}

\begin{figure}
\includegraphics[scale=0.5]{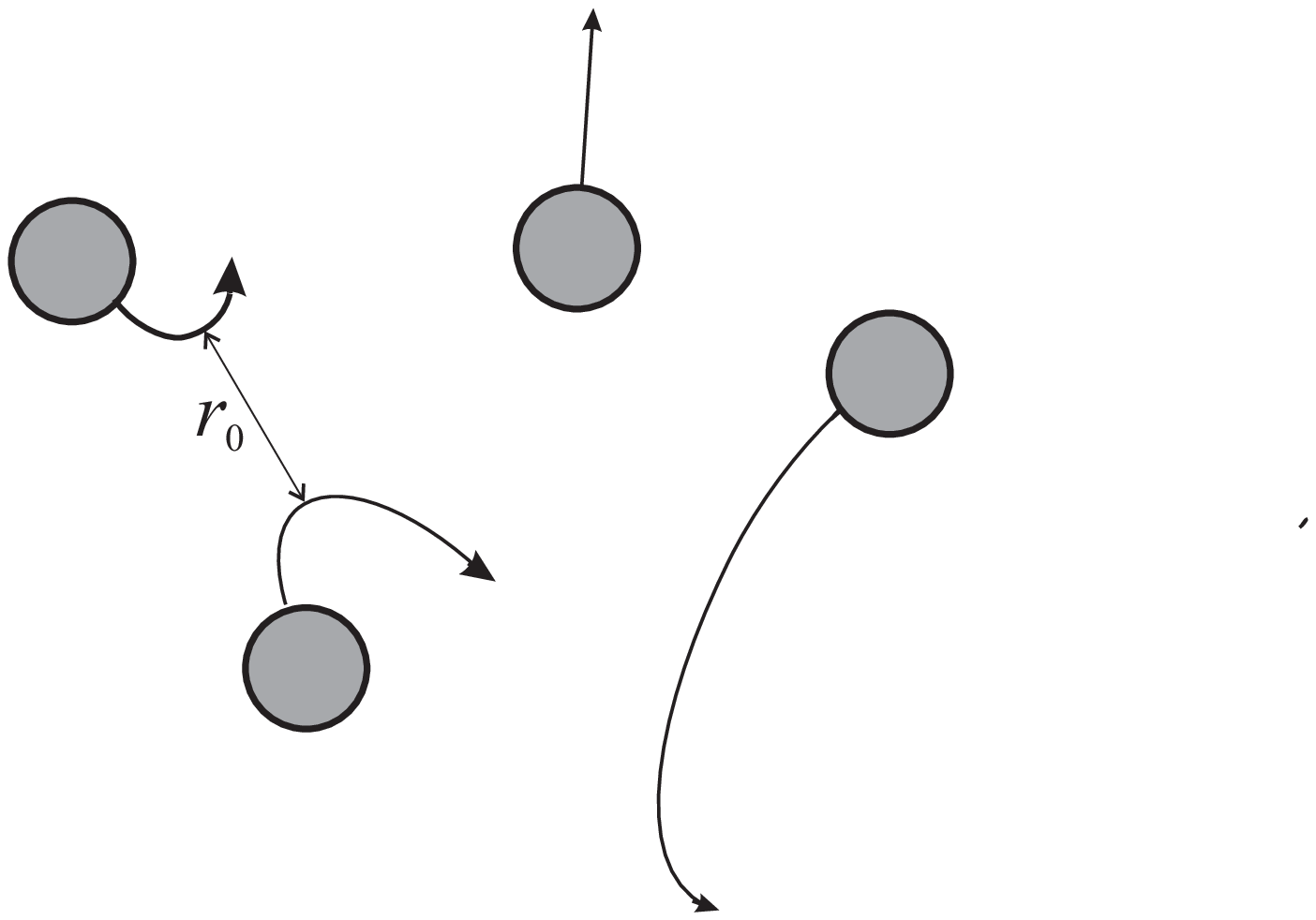}
\caption{\label{fig:Classical_gas}At low temperatures, $T<e^{2}/\kappa d$, the
exciton kinetic energy cannot overcome the repulsion between the excitons. The
minimal distance between excitons in average is larger than $r_{0}$ defined in
Eq.(\ref{eq:mfa.4}) where the energy $E\sim T$. The correlations between
excitons cannot be neglected.}
\end{figure}

When the temperature goes down at some point it becomes smaller than the
Coulomb interaction at distance $d$ that violates the condition of the mean
field approximation, Eq.(\ref{eq:mfa.5}), and the correlation of excitons
cannot be neglected anymore, Fig.\ref{fig:Classical_gas}. At $T<e^{2}/\kappa d$
the relevant parameter characterizing the scale of the interaction potential is
not $d$ but the minimal distance between excitons $r_{0}$ because $r_{0}>d$. At
the temperature where $r_{0}$ crosses the value of $d$, condition
(\ref{eq:mfa.3}) leads to
\begin{equation}
nr_{0}^{2} \ll 1 .
\label{eq:cxg.1}
\end{equation}
This means that right below the temperature $T\sim e^{2}/\kappa d$ there exists
a region where the average distance between excitons is larger than $r_{0}$ and
it is possible to take into account only pair correlations because the
probability to find three excitons in mutual proximity is negligible. The
condition $r_{0}\sim d$ (i.e., $T\sim e^{2}/\kappa d$) means also that
$k_{T}r_{0}\sim k_{T}d\gg k_{T}\sqrt{db}\sim1$ (the last two relations follow
from $d\gg b$ and $T\sim e^{2}/\kappa d$ respectively) and the interaction
between excitons can be still considered classically. The same relations mean
that $k_{T}^{2}\gg n$, i.e., the exciton gas is non-degenerate. A growth of
$r_{0}$ compared to the exciton radius also makes the exchange corrections to
the interaction (\ref{eq:mfa.1}) less important. As a result at the temperature
region right below $e^{2}/\kappa d$ it is possible to consider excitons as
classical particles.

\begin{figure}
\includegraphics[scale=0.5]{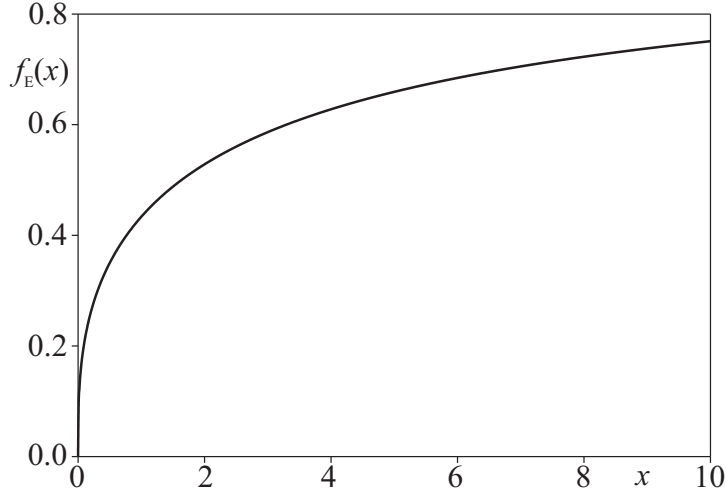}
\caption{\label{fig:fE}A plot of dimensionless function $f_{E}(x)$ of Eq.
(\ref{eq:cxg.5}) that defines the blue shift at low exciton concentrations,
Eq.(\ref{eq:cxg.1}), and moderate temperatures, Eq.(\ref{eq:cxg.14}).}
\end{figure}

The blue shift in this temperature and concentration region can be evaluated as
\begin{equation}
E_{int} = n \int U(r)g(r) d^{2}\bm{r} \ ,
\label{eq:cxg.2}
\end{equation}
where $g(r)$ is the pair correlation function. For any given exciton,
$ng(r)d^{2}\bm{r}$ is the average number of excitons within area $d^{2}\bm{r}$
at distance $r$ from it. According to this definition
$g(r)|_{r\rightarrow\infty}=1$ because at large distance any correlation
between excitons disappears. In the leading order in $nr_{0}^{2}$
\begin{equation}
g(r) = e^{-U(r)/T}
\label{eq:cxg.3}
\end{equation}
[see, e.g., Ref.\cite{Hill} Sec.32]. Substitution of Eq.(\ref{eq:cxg.3}) into
Eq.(\ref{eq:cxg.2}) gives
\begin{equation}
E_{int} = n \int U(r)e^{-U(r)/T} d^{2}\bm{r} \ .
\label{eq:cxg.4}
\end{equation}
With the interaction Eq.(\ref{eq:mfa.1}) this expression is reduced to a
function of only one parameter,
\begin{equation}
E_{int} = \frac{4\pi ne^{2}d}{\kappa} \
    f_{E}\left(\frac{\kappa dT}{e^{2}}\right) .
\label{eq:cxg.5}
\end{equation}
Function $f_{E}(x)$ is plotted in Fig.\ref{fig:fE}.
The asymptotical behavior of this function is
\begin{equation}
f_{E}(x) =  \left\{
\begin{array}{ll}
    \displaystyle
    1 + \frac{1}{x} \left(\ln\frac{8}{x} + C - 1\right) \ , \hspace{2cm}
    & x \gg 1 \ , \\[0.1cm]
    \displaystyle \frac{\Gamma(4/3)}{2} \ x^{1/3} \ , & x \ll 1 \ ,
\end{array} \right.
\label{eq:cxg.6}
\end{equation}
where $C\approx0.577$ is the Euler constant. At high temperature,
Eq.(\ref{eq:mfa.5}), the expression for $E_{int}$ (\ref{eq:cxg.5}) is reduced
to the mean field expression of Eq.(\ref{eq:mfa.2}). In the opposite limit,
where $r_{0}\gg d$, the interaction potential is simplified:
\begin{equation}
U(r) = \frac{e^{2}d^{2}}{\kappa r^{3}} \ , \hspace{1cm} r \gg d \ ,
\label{eq:cxg.7}
\end{equation}
and Eq.(\ref{eq:cxg.4}) gives
\begin{equation}
E_{int} = 2\pi\Gamma(4/3)
    n \left(\frac{e^{2}d^{2}}{\kappa}\right)^{2/3}T^{1/3} =
    2\pi\Gamma(4/3)nr_{0}^{2}T \ ,
\label{eq:cxg.8}
\end{equation}
where $r_{0}=(e^{2}d^{2}/\kappa T)^{1/3}$. Qualitatively this result can be
understood in the following way. Around each exciton there is a depletion
region with a radius $\sim r_{0}$. Without repulsion this region would contain
$nr_{0}^{2}$ excitons with average energy $T$. The energy necessary to force
all of them out of the region is of the order of $nr_{0}^{2}T$.

One can expect that $E_{int}$ is of the order of the first virial correction to
the chemical potential of the exciton gas. This can be easily checked. The
first two terms of the virial expansion give (Ref.\cite{Hill} Sec.23)
\begin{equation}
n = \frac{2mT}{\pi\hbar^{2}} \ e^{\zeta/T} +
    \left(\frac{2mT}{\pi\hbar^{2}} \ e^{\zeta/T}\right)^{2}
    \int \left(e^{-U(r)/T} -  1\right) d^{2}\bm{r} \ .
\label{eq:cxg.9}
\end{equation}
Solution of this equation with respect to the chemical potential $\zeta$ leads
to
\begin{equation}
\zeta = \zeta_{0} + \Delta\zeta \ , \hspace{1cm}
    \zeta_{0} = T \ln\frac{\pi\hbar^{2}n}{2MT} \ ,
\label{eq:cxg.10}
\end{equation}
and
\begin{subequations}
\begin{eqnarray}
&& \Delta\zeta = - nT \int \left[e^{-U(r)/T} - 1\right] d^{2}\bm{r} =
    \frac{2\pi ne^{2}d}{\kappa} \
    f_{\zeta}\left(\frac{\kappa dT}{e^{2}}\right) \ ,
\label{eq:cxg.11a} \\
&& f_{\zeta}(x) = x \int_{0}^{\infty}
        \left\{
    1 - \exp\left[
    - \frac{2}{x} \left(\frac{1}{t} - \frac{1}{\sqrt{t^{2} + 1}}\right)
            \right]
        \right\} t dt \ .
\label{eq:cxg.11b}
\end{eqnarray}
\label{eq:cxg.11}
\end{subequations}
In extreme cases
\begin{equation}
\Delta\zeta =   \left\{
\begin{array}{ll}
    \displaystyle   \frac{4\pi ne^{2}d}{\kappa} \ , &
    \displaystyle   T \gg \frac{e^{2}}{\kappa d} \ , \\
    \displaystyle
    \pi\Gamma(1/3) \left(\frac{e^{2}d^{2}}{\kappa}\right)^{2/3} nT^{1/3} \ , &
    \displaystyle \hspace{1cm} T \ll \frac{e^{2}}{\kappa d} \ .
\end{array}     \right.
\label{eq:cxg.12}
\end{equation}
That is $E_{int}$ and $\Delta\zeta$ have the same dependence on parameters and
differ only by a constant factor. This difference results from the difference
of the definitions: $E_{int}$ is the correction to the average energy released
in an exciton recombination while $\Delta\zeta$ is the correction to the energy
released when an exciton is removed without a violation of the equilibrium in the
exciton gas.

Finally, we note that in  the case of $T\ll e^{2}/\kappa d$ when $E_{int}$ is
given by Eq.(\ref{eq:cxg.8}), the interaction at the average distance
$U(n^{-1/2})$ is still small compared to $E_{int}$: $U(n^{-1/2})/E_{int}\sim
n^{1/2}r_{0}\ll1$. The mean field approximation then gives correct expression
only for the average energy because
\begin{equation}
\overline{U^{2}} = n \int U^{2}(r)e^{-U(r)/T} d^{2}\bm{r} =
    \frac{2\pi\Gamma(4/3)}{3} \ nr_{0}^{2}T^{2} \sim
    E_{int}^{2}/nr_{0}^{2} \ ,
\label{eq:cxg.13}
\end{equation}
i.e., the fluctuations of the interaction energy are larger than its average value.

The results of this section, Eqs.(\ref{eq:cxg.4}) and (\ref{eq:cxg.8}), are
valid under two conditions: that of small concentration, Eq.(\ref{eq:cxg.1}),
and
\begin{equation}
k_{T}r_{0} = \left(\frac{2d^{2}k_{T}}{b}\right)^{1/3} \gg 1 \ ,
\label{eq:cxg.14}
\end{equation}
that validates classical description of the interaction.

The parameter $nr_{0}^{2}$ is the gas parameter which is the ratio of the
exciton interaction energy to its kinetic energy. The same parameter indicates
the strength of the exciton -- exciton scattering. If the impact parameter in a
scattering event of two excitons is $\lesssim r_{0}$ then the scattering angle
is large. Therefore the scattering cross-section (in 2D case it has units of
length) is $\sim r_{0}$. Respectively, the mean free path is $l\sim1/nr_{0}$.
The scattering is weak, i.e., three and more particle scattering can be
neglected if $l$ is larger than the interparticle distance which means that the
gas parameter is small, Eq.(\ref{eq:cxg.1}). If this condition is violated the
exciton system cannot be considered as a gas, it is a liquid. On the other
hand, parameter $k_{T}r_{0}$ characterizes quantum corrections to scattering of
excitons. Eq.(\ref{eq:cxg.14}) is stronger than the non-degeneracy condition
because $k_{T}r_{0}=(k_{T}/n^{1/2})(n^{1/2}r_{0})\ll k_{T}/n^{1/2}$ because the
characteristic scale of the potential $r_{0}$ is smaller than the average
distance between excitons, Eq.(\ref{eq:cxg.1}).

When temperature goes down further both condition of small concentration,
Eqs.(\ref{eq:cxg.1}), and classical description of exciton - exciton
scattering, Eq.(\ref{eq:cxg.14}), are at some point violated. Which one is
violated first depends on the concentration. If $n<(b/2d^{2})^{2}$ then quantum
effects in the exciton scattering become important when the exciton system can
still be considered as a non-degenerate gas. In the opposite case with
reduction of the temperature the exciton system becomes a liquid before any
quantum corrections to the scattering process become pronounced.

\section{Exciton gas with quantum scattering}
\label{sec:xgqs}

If
\begin{equation}
n \ll (b/2d^{2})^{2}
\label{eq:xgqs.1}
\end{equation}
and $k_{T}r_{0}\lesssim1$ which is equivalent to
\begin{equation}
k_{T} \lesssim b/2d^{2},
\label{eq:xgqs.2}
\end{equation}
then quantum corrections to the exciton -- exciton scattering are important but the
exciton gas is yet non-degenerate until $k_{T}\gg n^{1/2}$. For the calculation of the
interaction energy it is then still possible to use Eq.(\ref{eq:cxg.2}) but $g(r)$ has to
be modified to include quantum corrections. This can be done in the following
way.

A wave function describing a state of two excitons can be factorized into a
wave function of the center of mass and a wave function $\psi(\bm{r})$
describing their relative motion (see Appendix \ref{sec:xxs}). $\psi(\bm{r})$
is characterized by a few quantum numbers but in equilibrium the occupation of
a state depends only on its energy. This means that, given the energy of the
relative motion $E$ of any two excitons, the probability to find one exciton at
the distance $r$ from the other is $\langle|\psi(\bm{r})|^{2}\rangle_{E}$ where
$\langle\dots\rangle_{E}$ is the average over all quantum numbers (e.g., the
direction of the wave vector) except $E$. For a non-degenerate exciton gas the
probability density for an exciton to have energy $E$ is $(1/T)e^{-E/T}$. That
is
\begin{equation}
g(r) = \frac{1}{T}
    \int_{0}^{\infty} \langle|\psi(\bm{r})|^{2}\rangle_{E} e^{-E/T} dE
\label{eq:xgqs.3}
\end{equation}
where $\psi(\bm{r})$ has to be normalized in such a way that
$g(r)|_{r\rightarrow\infty}=1$ which corresponds to usual normalization for
scattering problem.

In a semiclassical approximation, when the exciton wavelength is smaller than the
length scale of the interaction potential, $k_{T}r_{0}\gg1$,
\begin{equation}
\psi(\bm{r}) = \frac{A}{[E - U(r)]^{1/4}} \ e^{iS/\hbar} \ ,
\label{eq:xgqs.4}
\end{equation}
where $S$ satisfies the equation
\begin{equation}
\frac{(\nabla S)^{2}}{M} = E - U(r) \ .
\label{eq:xgqs.5}
\end{equation}
Substitution of Eq.(\ref{eq:xgqs.4}) in Eq.(\ref{eq:xgqs.3}) results in
\begin{equation}
g(r) = \frac{A^{2}}{T}
    \int_{U(r)}^{\infty} e^{-E/T} \ \frac{dE}{\sqrt{E - U(r)}} =
    \frac{A^{2}\sqrt{\pi}}{\sqrt{T}} \ e^{-U(r)/T}
\label{eq:xgqs.6}
\end{equation}
which is identical to Eq.(\ref{eq:cxg.3}) for $A^{2}=\sqrt{T/\pi}$. Fig.11 of
Ref.\cite{Zimmermann07} for $g(r)$ is related to an intermediate case where
$k_{T}r_{0}\approx1.7$ and it differs from our classical expression by 20\% in
the scale of $r$ which comes from not very large value of $k_{T}r_{0}$ and our
simplification of the interaction between excitons.

\begin{figure}
\includegraphics[scale=0.5]{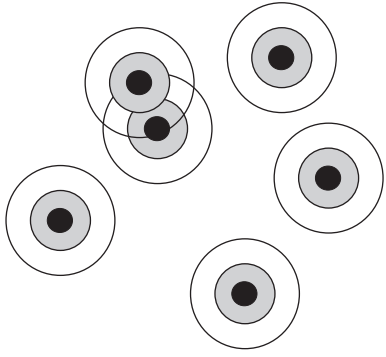}
\caption{\label{fig:Quantum_gas}A sketch of an exciton gas with quantum
scattering. The radius of the white circle is of the order of the exciton
wavelength, $\sim1/k_{T}$, the radius of the gray circle is the classical
minimal distance between excitons, $\sim r_{0}$, and the radius of the black
circle is the quantum minimal distance between excitons, $\sim d^{2}/b$. Here
the exciton wavelength is larger than $r_{0}$ and the scattering is essentially
quantum mechanical. On the other hand the average distance between excitons is
larger than their wavelength so that the gas is non-degenerate.}
\end{figure}

If $k_{T}r_{0}\ll1$ the result strongly differs from the classical case. The
wave function $\psi(\bm{r})$ penetrates under the repulsion barrier and the
minimal distance between excitons is characterized not by $r_{0}$ anymore but
rather by the distance at which $\psi(\bm{r})$ falls off. The dipole -- dipole
repulsion $e^{2}d^{2}/\kappa r^{3}$ is trying to push the wave function to
larger $r$ while the kinetic energy $\sim(\hbar^{2}/Mr^{2})$ is trying to
spread it to all available space and in particular to as small values of $r$ as
possible. The distance at which $\psi(\bm{r})$ falls off is characterized by
the same order of magnitude of these two tendencies. This gives the distance of
$\sim d^{2}/b$, Fig.\ref{fig:Quantum_gas}.

When the interaction potential is approximated by interaction between point
like dipoles, Eq.(\ref{eq:cxg.7}), this distance is smaller than $r_{0}$:
$r_{0}/(d^{2}/b)\sim(b/k_{T}d^{2})^{2/3}\gg1$ because of Eq.(\ref{eq:xgqs.2}).
But due to $b\ll d$ it is much larger than $d$ which justifies the
approximation of point-like dipoles, Eq.(\ref{eq:cxg.7}).

The interaction energy can be estimated as $U(d^{2}/b)$ (under the condition
$k_{T}r_{0}\ll1$ it is larger than $T$) multiplied by the number of excitons in
the important region, $n(d^{2}/b)^{2}$, that gives $E_{int}\sim\hbar^{2}n/m$.
The exact calculation (Appendix \ref{sec:xxs}) shows that $\psi(\bm{r})$ falls
off exponentially when $r<d^{2}/b$ and is a relatively slow function of $r$
when $r>d^{2}/b$. On the other hand, the potential $U(r)$ falls off as
$1/r^{3}$ when $r>d^{2}/b$. As a result, the main contribution to the
interaction energy Eq.(\ref{eq:cxg.2}) comes from the region $r\sim d^{2}/b$.
In this region the wave function is given by Eqs.(\ref{eq:xxs.7}) and
(\ref{eq:xxs.12}):
\begin{equation}
\psi(r) = - \frac{2}{\ln(kd^{2}/b)} \
    K_{0}\left(\frac{2d}{\sqrt{br}}\right) ,
\label{eq:xgqs.7}
\end{equation}
where $k=\sqrt{ME}/\hbar$. The logarithmic dependence of this wave function on
the energy is very weak and with a logarithmic accuracy the substitution of
Eq.(\ref{eq:xgqs.7}) into Eq.(\ref{eq:xgqs.3}) leads to the following expression
for the correlation function:
\begin{equation}
g(r) = \frac{4n}{\ln^{2}(k_{T}d^{2}/b)} \
    K_{0}^{2}\left(\frac{2d}{\sqrt{br}}\right) , \hspace{1cm} r \ll 1/k_{T} \ .
\label{eq:xgqs.8}
\end{equation}
As a result,
\begin{equation}
E_{int} = \frac{8\pi n}{\ln^{2}(k_{T}d^{2}/b)} \
    \frac{e^{2}d^{2}}{\kappa}
    \int_{0}^{\infty} K_{0}^{2}\left(\frac{2d}{\sqrt{br}}\right)
    \frac{dr}{r^{2}} =
    \frac{2\pi\hbar^{2}n}{M\ln^{2}(k_{T}d^{2}/b)} \ ,
\label{eq:xgqs.9}
\end{equation}
where Eq.(6.576.4) of Ref.\cite{gradsteyn} has been used. Note that in this
case again $E_{int}$ is larger than the interaction at the average distance
between excitons:
$U(n^{-1/2})/E_{int}\sim(d^{2}n^{1/2}/b)\ln^{2}(k_{T}d^{2}/b)\ll1$ which is due
to Eq.(\ref{eq:xgqs.1}).

The gas parameter in the quantum case is different from classical one,
$nr_{0}^{2}$. Exciton -- exciton scattering is strong if two excitons approach
each other to a distance equal to their wavelength. In other words, the
scattering cross-section is of the order of the wavelength with accuracy of a
logarithmic correction (this is the well known difference between the 2D and
the 3D case where at small wave vectors the cross-section goes to a constant,
see Ref.\cite{landau_qm}, Problem 7 to Sec.132, and Appendix \ref{sec:xxs}) in
spite of the fact that the length scale of the potential $d^{2}/b$ is smaller
than the wavelength. Respectively the mean free path of excitons is $l\sim
k_{T}/n$. The gas condition in quantum case is the absence of correlations
between different scattering events which means that the wavelength has to be
much smaller than the mean free path, i.e., $k_{T}l\gg1$ or
\begin{equation}
n \ll k_{T}^{2}\ln^{2}(kd^{2}/b) \ .
\label{eq:xgqs.10}
\end{equation}
This inequality is identical with $E_{int}/T\ll1$ and also with the condition
of non-degeneracy with an accuracy of the logarithmic correction. Practically the
logarithm is not very large and \textit{in the gas state the exciton system is
non-degenerate while the degeneracy is accompanied with strong interactions and
multi-particle correlations between the excitons which is characteristic for liquids}.

\section{Liquid state of the exciton system}
\label{sec:lsxs}

\begin{figure}
\includegraphics[scale=0.5]{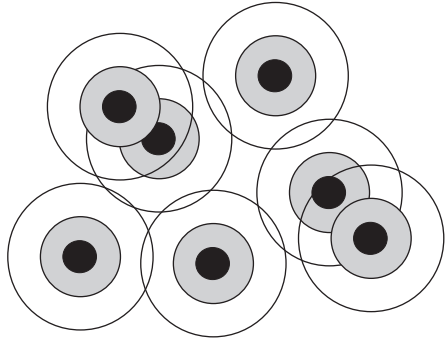}
\caption{\label{fig:Quantum_liquid}A rough sketch of quantum exciton liquid.
The radius of the white circle is of the order of the exciton wavelength,
$\sim1/k_{T}$, the radius of the gray circle is the classical minimal distance
between excitons, $\sim r_{0}$, the radius of the black circle is the quantum
minimal distance between excitons, $\sim d^{2}/b$. The average distance between
excitons is of the order or smaller than their wavelength and the system is
degenerate. But the same condition means that multi-particle correlation is
strong and this is a liquid. The picture cannot demonstrate that different
excitons are not distinguishable due to an overlap of their wave functions.}
\end{figure}

In this section we consider the temperatures and/or concentrations beyond the
limits specified in previous sections. In those cases multiexciton correlations
are important and the problem is not reduced to a two-particle problem,
Fig.\ref{fig:Quantum_liquid}. Actually all close neighbors are correlated
although a long range correlation may not exist. This situation is
characteristic for liquids and therefore we use the term "liquid" for such
states of exciton system.

An exact microscopical theory of liquids does not exist and to avoid poorly
controllable and rather complicated approximations we make estimates based on
reasonable physical arguments. These estimates give not only an approximate
value of the blue shift but also its dependence on the concentration and
temperature. Also, the arguments and estimates allow us to develop a general
understanding of the structure of the exciton system at low temperatures and
high concentrations.

The picture is different depending on the exciton concentration compared to
$(b/2d^{2})^{2}$.

\subsection{Low concentration}

If the concentration is low according to quantum criterion
Eq.(\ref{eq:xgqs.1}), and the temperature is low enough so that multi-particle
correlations take place, i.e., the inequality in Eq.(\ref{eq:xgqs.10}) is
violated, then Eq.(\ref{eq:xgqs.9}) can be considered as a good estimate that
can have only logarithmic corrections. Indeed, the temperature in
Eq.(\ref{eq:xgqs.9}) enters only in the argument of the logarithm and only this
argument can change when the temperature goes down. We emphasize once again
that violation of condition (\ref{eq:xgqs.10}) leads not only to degeneracy but
also to multi-particle correlation which makes the theory of dilute Bose gas
unapplicable.

We would like to attract an attention to a generally known fact that in a dilute
2D Bose gas the characteristic energy at low temperatures does not depend on the
coupling constant, except logarithmic corrections. This comes from the virial
theorem, i.e., from comparison of the interaction energy and kinetic energy and
is a generalization of Eq.(\ref{eq:xgqs.9}) to any interaction between
particles. In a dilute gas only two particles can be at the distance where
their interaction is important (large quantum uncertainty of the distance
$\sim1/k_{T}$ means only that we cannot be sure that they are at this
distance). If the interaction between particles is $U(r)$ then from the virial
theorem it follows that $U(r)\sim\hbar^{2}/Mr^{2}$. The value of $r$ obtained
from this relation is the distance at which the interaction is important. The
interaction energy of a particle is the interaction energy between two
particles at the distance $\lesssim r$ times the probability that two particles
come to this distance, $nr^{2}$. This gives $n\hbar^{2}/M$. In a liquid the
estimate can have a numerical factor characterizing the number of particles
within the interaction radius. This energy gives a temperature scale for both
Bose condensation\cite{Popov,Fisher88} and Kosterlitz -- Thouless
transition\cite{Lozovik96,Koinov00}.

\subsection{High concentration}

If the concentration is beyond its quantum limitation, i.e.,
Eq.(\ref{eq:xgqs.1}) is not satisfied then decrease of the temperature or
increase of the concentration leads to violation of the classical gas
condition, Eq.(\ref{eq:cxg.1}), while the system is still non-degenerate, i.e.,
Eq.(\ref{eq:cxg.14}) holds. In this case the exciton system becomes a classical
liquid. As long as Eq.(\ref{eq:cxg.7}) is valid the dimensional analysis allows
us to express $E_{int}$ as a function of only one parameter:
\begin{equation}
E_{int} = \frac{e^{2}d^{2}}{\kappa} \ n^{3/2}
    f\left(\frac{e^{2}d^{2}}{\kappa T} \ n^{3/2}\right) \ .
\label{eq:lsxs.hc.1}
\end{equation}
According to Eq.(\ref{eq:cxg.8}) $f(x)=2\pi\Gamma(4/3)x^{-1/3}$ at $x\ll1$.

\begin{figure}
\includegraphics[scale=0.4]{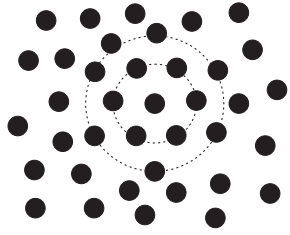}
\caption{\label{fig:Class_liquid}A rough sketch of a short range order in
a classical liquid. In this example the order extends to two coordinate circles, and at larger
distance excitons are not correlated. (Note that this picture is oversimplified: similar
order exists around most of the excitons).}
\end{figure}

When $(e^{2}d^{2}/\kappa T)n^{3/2}=r_{0}^{3}n^{3/2}$ grows and becomes of the
order of unity a free motion of excitons between collisions becomes impossible
because each of them is confined in between its neighbors. In other words an
exciton is in a highly excited state in a potential well formed by its
neighbors. The size of the well is $R\sim n^{-1/2}$ and this semi-classical
picture is valid as long as the size of the confinement region is much larger
than the exciton thermal wavelength, i.e. $k_{T}R\gg1$. The energy at the
bottom of the potential well is $\sim ze^{2}d^{2}/R^{3}$ ($z$ is the number of
nearest neighbors) and is of the same order as the depth of the well. Potential
wells for different excitons are different, they are not static and sometimes
some excitons overcome or tunnel across the surrounding barrier. But at
$nr_{0}^{2}>1$ these rear occasions do not affect the estimates. In general, this
picture is similar to a simple classical liquid and the formation of the potential
wells is the starting point of the formation of a short range order characteristic
for liquids.\cite{March} We emphasize that we mean a formation of a short range
order typical in liquids but not crystallization and a formation of a long range
order\cite{Lozovik98}, Fig.\ref{fig:Class_liquid}. Further reduction of the
temperature brings particles to lower levels in the potential wells and makes
the wells more stable. A stronger confinement of the wave functions of each
exciton reduces their overlap. When the size of the exciton wave function
becomes smaller than $R$, the potential for each exciton can be approximated as
\begin{equation}
U_{liq}(\bm{r}) \approx \frac{e^{2}d^{2}}{\kappa R^{3}}
    \left(C_{1} + C_{2} \ \frac{r^{2}}{R^{2}}\right) ,
\label{eq:lsxs.hc.2}
\end{equation}
where $r$ is the distance from the minimum of the well. We estimate the constants
$C_{1}$ and $C_{2}$ assuming a short range order, i.e., there is a crystal
structure around an exciton within $l$ coordinate circles but beyond this
region the exciton positions are not correlated. This assumption gives
\begin{subequations}
\begin{eqnarray}
C_{1} & = & \sum_{j=1}^{l} \frac{z_{j}}{(R_{j}/R)^{3}} +
    \frac{2\pi}{R_{l+1}/R} \ ,
\label{eq:lsxs.hc.3a} \\
C_{2} & = & \frac{9}{4}
    \sum_{j=1}^{l} \frac{z_{j}}{(R_{j}/R)^{5}} +
    \frac{3\pi}{2(R_{l+1}/R)^{3}} \ .
\label{eq:lsxs.hc.3b}
\end{eqnarray}
\label{eq:lsxs.hc.3}
\end{subequations}
Here $R=[n\sin(2\pi/z)]^{-1/2}$ is the lattice constant which is the same as
the radius of the first coordinate circle, $R_{j}$ is the radius of the $j$th
coordinate circle, and $z_{j}$ is the number of particles at the $j$th
coordinate circle, $z_{1}=z$. The resulting values of the constants appear to
be very weakly sensitive to the radius of the order, see Table \ref{tab:const}
(compare Ref.\cite{Rapaport07}).

\begin{table}
\caption{\label{tab:const}Constants which characterize the effective potential well for each
exciton in a liquid, Eq.(\ref{eq:lsxs.hc.2}). $l=0$ corresponds to very small
correlations between the exciton positions, and $l=\infty$ corresponds to a crystal
structure.} \centering
\begin{tabular}{||c||c|c|c|c||}
\hline
  Coordinate circle  & \multicolumn{2}{|c|}{Square lattice, $z=4$} &
            \multicolumn{2}{|c||}{Hexagonal lattice, $z=6$}          \\
\hline
    $l$   & $C_{1}$ & $C_{2}$    & $C_{1}$ & $C_{2}$            \\
\hline
    0     & 6.28    & 4.71       & 7.25    & 5.44               \\
    1     & 8.44    & 10.67      & 10.19   & 14.55              \\
    2     & 8.55    & 11.18      & 10.78   & 15.07              \\
    3     & 8.72    & 11.29      & 10.65   & 15.08              \\
 $\infty$ & 8.84    & 11.37      & 10.84   & 15.13              \\
\hline
\end{tabular}
\end{table}

If $T\sim(e^{2}d^{2}/\kappa)n^{3/2}$ (i.e., $nr_{0}^{2}\sim1$) then $r\sim R$
and there is no short range order in the system. However, if $T
\ll(e^{2}d^{2}/\kappa)n^{3/2}$ then the short range order does exist, and most
of the excitons are at the ground state of their corresponding potential
Eq.(\ref{eq:lsxs.hc.2}). The energy of the ground state above the bottom of the
potential well and its radius are
\begin{equation}
\hbar\omega_{gs} \approx
    \sqrt{2C_{2} \ \frac{\hbar^{2}}{MR^{2}} \ \frac{e^{2}d^{2}}{\kappa R^{3}}}
    = \frac{2C_{2} \ \hbar^{2}n}{M} \sqrt{\frac{d^{2}n^{1/2}}{b}} \ ,
    \hspace{1cm}
    r_{gs} = R \left(\frac{1}{2C_{2}} \ \frac{bR}{d^{2}}\right)^{1/4} \ .
\label{eq:lsxs.hc.4}
\end{equation}

The characteristic size of the exciton wave function $r$ is controlled by the
temperature or the energy of the ground state, whichever is larger, and in any
case when $T\ll(e^{2}d^{2}/\kappa)n^{3/2}$ \textit{the size of a single exciton
wave function is much smaller than the distance between the excitons}.

The inequality $r\ll R$ allows us to make two conclusions. First, the bottom of
potential $U_{liq}(\bm{r})$, Eq.(\ref{eq:lsxs.hc.2}), gives a good estimate for
the interaction energy
\begin{equation}
E_{int} \approx \frac{10e^{2}d^{2}n^{3/2}}{\kappa} \ .
\label{eq:lsxs.hc.5}
\end{equation}
Comparison of this expression with Eq.(\ref{eq:lsxs.hc.1}) shows that
$f(x\rightarrow\infty)\approx10$. This estimate does not include possible
logarithmic corrections.

Second, it is possible to estimate the overlap of the wave functions of
adjacent excitons. If $d=12$ nm and $n=2\times10^{11}$ cm$^{-2}$ the estimate
according to the wave function in the harmonic potential of
Eq.(\ref{eq:lsxs.hc.2}) gives for the overlap a value of $\sim0.14$. The actual
value is even smaller because at $r\sim R$ the potential barrier is steeper
than the harmonic one. Due to the small wave function overlap the temperature
at which the phase or/and spin coherence\cite{Fernandez-Rossier96} in the
exciton system is set in is reduced compared to its the expected value
$\sim\hbar^{2}n/M$. This points to a possible non-monotonic dependence of the
quantum coherence onset temperature on the concentration, and it suggests that
a lower density exciton system may become quantum coherent at higher
temperatures than a higher density system, which is a-priori non intuitive. In
other words, a long range interaction suppresses quantum degeneracy.

At low temperature the physics of the transition between quantum liquid and
classical liquid with growth of the exciton concentration is the following. At
low concentration $n\ll(b/d^{2})^{2}$, Eq.(\ref{eq:xgqs.1}), according to the
exciton wave function Eq.(\ref{eq:xgqs.7}) around each exciton there is a
circle with radius $\sim d^{2}/b$ inside which the wave function of any other
exciton is exponentially small. However, the radius of this circle is much
smaller than the average distance between excitons, $n^{-1/2}$, and any exciton wave function can easily spread at the area that contains many other excitons avoiding their depleted circular regions. That is wave functions of different
excitons overlap forming a quantum liquid (not a gas because of strong exciton -- exciton scattering).\footnote{Strictly speaking, we have to talk about
multi-particle wave function $\psi(\bm{r}_{1},\bm{r}_{2},\dots)$ that goes to
zero when any two of their arguments become closer than $d^{2}/b$ and is more
or less the same order of magnitude at all other values of the arguments.} With growth of the concentration the average distance between excitons decreases
that makes spreading of each exciton wave function to a wide area more difficult. This reduces the overlap of the wave functions of different excitons. Finally, when the average distance becomes smaller than the radius of the circular depleted region, $n^{-1/2}\lesssim b/d^{2}$, the wave function of nearly each exciton appears to be confined in between its nearest neighbors. The overlap of the wave functions of adjacent excitons is very weak and the system becomes a classical liquid.

\section{Discussion}

\begin{figure}
\includegraphics[scale=0.4]{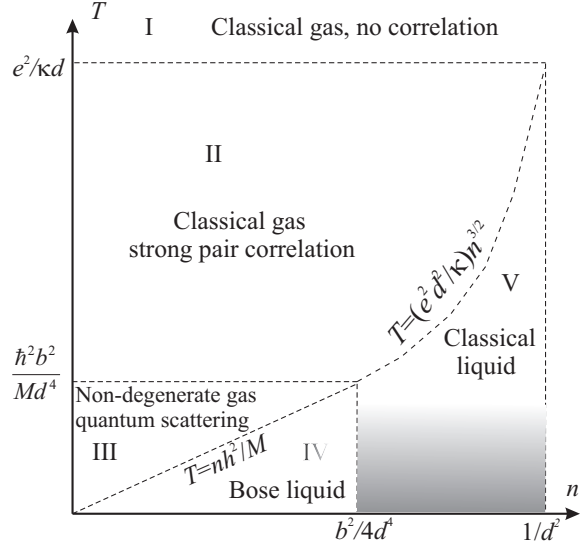}
\caption{\label{fig:nT}Different role of the exciton - exciton correlations in
an exciton system. The shaded area shows the part of the region V where quantum
correlations set in.}
\end{figure}

Estimates made in the previous sections open the possibility to develop a
general picture that demonstrates a role of correlations in the exciton system
at the whole $n-T$ plane. This picture is presented in Fig. \ref{fig:nT}.
Correlations are not important and the mean field approximation is applicable
only in region I. In region II the exciton system can be considered as a
classical gas with strong pair correlations. In region III, contrary to region
II, the exciton - exciton scattering is described by quantum mechanics. In
other respects this region is similar to region II. Reduction of the
temperature from region III to region IV leads to degeneracy of the exciton
system. But simultaneously a strong multi - particle correlation is set up. The
system cannot be considered as a dilute gas, and the mean free path does not
exist. Rather surprising is the existence of region V where the system behaves
as a classical liquid down to temperatures well below $\hbar^{2}n/M$ (compare
Ref.\cite{Buchler07}). The reason is that strong repulsion between excitons
squeezes the wave function of each exciton to an area smaller than the average
average area per one exciton. In this region a short range order appears and
with further reduction of temperature its correlation radius grows. However,
contrary to regular classical liquids, the attractive part of the exciton -
exciton interaction is negligible \cite{Lozovik96,Zimmermann07} and it is
likely that a long range order is settled not as a result of a phase transition
but as gradual growth of the correlation radius.

A comparison the of values of the exciton binding energy $\epsilon_{b}$ and the
Coulomb interaction at distance $d$, $e^{2}/\kappa d$, in Table \ref{tab:param}
leads to the conclusion that the correlations between excitons can be neglected
only when a significant part of them is dissociated. However, in this case the
concentration $n$ that controls the blue shift in Eq.(\ref{eq:1}) is not the
exciton concentration but the sum of the exciton concentration and
concentration of electrons or holes.

The existence of region V is the result of the tail of the exciton - exciton
repulsion potential. In case of a short range potential (e.g., hard circles) an
overlap of the particle wave functions competes with the repulsion and the region
of classical liquid behavior does not exist. This happens to excitons in one well
where there is no dipole - dipole repulsion and an increase of the concentration
leads to the Mott transition but not to a classical liquid.

\begin{table}
\caption{\label{tab:param}Estimates for the parameters of Fig.\ref{fig:nT} for
GaAs/AlGaAs quantum well structures with different values of $L$.} \centering
\begin{tabular}{|c||c|c|c|c|c|}
\hline
 $d$ (nm)  & $\epsilon_{b}$ (K) & $e^{2}/\kappa d$ (K)
    & $\hbar^{2}b^{2}/Md^{4}$ (K) & $b^{2}/4d^{4}$ (cm$^{-2})$
        & $1/d^{2}$ (cm$^{-2})$ \\
\hline
  10 & 54. & 139  &  3.8  & $2.2\times10^{10}$  & $1.0\times10^{12}$  \\
  12 & 49. & 116  &  1.8  & $1.1\times10^{10}$  & $0.69\times10^{12}$ \\
  14 & 44. &  99  &  1.0  & $0.58\times10^{10}$ & $0.51\times10^{12}$ \\
\hline
\end{tabular}
\end{table}

Two of the lines in Fig.\ref{fig:nT}, between regions III and IV and between
regions II and V, actually comprise one line at which the gas parameter
condition is violated. The gas parameter is the product of the concentration
and the scattering crosssection squared. Between regions II and V this
crosssection is classical, $\sim r_{0}$, while between regions III and IV it is
quantum, $\sim1/k_{T}$. The other two lines, between regions II and III and
between regions IV and V, separate classical and quantum interactions between
excitons. Also, it is necessary to note that at the bottom of region V some
quantum coherent phenomena are possible.

It is important to note that the lines separating different regions in
Fig.\ref{fig:nT} do not correspond to sharp transitions. Crossing of one of the
lines by changing the temperature or the concentration leads to a gradual
change of the correlation between excitons. Fig.\ref{fig:nT} demonstrates only
the role of correlation but not phases of the system.

\section{Conclusions}

We studied an exciton system in coupled quantum wells where electrons and
holes are confined in different wells and the main interaction between excitons
is a dipole -- dipole repulsion. We found that in the most part of the
temperature -- concentration plane the system is characterized by a strong
exciton -- exciton correlation. At some parts of this plane the system behaves
as a gas with a strong pair correlation. In other parts where the gas parameter condition
is violated, i.e., where the probability to find more than two excitons close to each
other becomes of the order of unity, the correlation is multi-excitonic and the
system has to be considered as a liquid. In particular, at low concentrations
degeneracy of the system is accompanied by a setting in of multi-exciton
correlations. At high concentration the strong confinement of each exciton wave
function due to repulsion between excitons suppresses quantum correlations. The blue
shift of the exciton luminescence has a different value and a different dependence
on the exciton temperature and concentration depending on how close excitons
can come to each other. Therefore it is a sensitive tool for measuring of the
exciton -- exciton correlations.

\section{Acknowledgements}

B.L. appreciates discussions with M. Stern and A. L. Efros.

\appendix

\section{Exciton - exciton scattering}
\label{sec:xxs}

Here the problem of the exciton - exciton scattering is considered under the
condition of small energy of relative motion,
\begin{subequations}
\begin{eqnarray}
E = \frac{\hbar^{2}k^{2}}{M} \ll \frac{e^{2}}{\kappa d} \ ,
\label{eq:xxs.1a}
\end{eqnarray}
and large wavelength,
\begin{equation}
k \ll b/2d^{2} \ .
\label{eq:xxs.1b}
\end{equation}
\label{eq:xxs.1}
\end{subequations}

For two excitons the center of mass momentum and the momentum of relative
motion are defined as
\begin{equation}
\bm{K} = \bm{k}_{1} + \bm{k}_{2} \ , \hspace{1cm}
    \bm{k} = \frac{\bm{k}_{1} - \bm{k}_{2}}{2} \ .
\label{eq:xxs.2}
\end{equation}

The two-exciton wave function is factorized
\begin{equation}
\Psi(\bm{r}_{1},\bm{r}_{2}) = \frac{1}{\sqrt{S}} \
    e^{i\bm{K}(\bm{r}_{1}+\bm{r}_{2})/2} \psi(\bm{r}_{1} - \bm{r}_{2}) \ ,
\label{eq:xxs.3}
\end{equation}
and the Schr\"odinger equation for the wave function describing their relative
motion is
\begin{equation}
- \frac{\hbar^{2}}{M} \ \nabla^{2}\psi(\bm{r}) + U(r)\psi(\bm{r}) =
    E\psi(\bm{r}) \ .
\label{eq:xxs.4}
\end{equation}

Under the condition Eq.(\ref{eq:xxs.1a}) the minimal distance between excitons is
much larger than $d$ (i.e., at $r\sim d$ the wave function is negligibly small)
and the interaction potential can be approximated with Eq.(\ref{eq:cxg.7}).
The condition of a very long exciton wavelength, Eq.(\ref{eq:xxs.1b}), makes it possible
to simplify Eq.(\ref{eq:xxs.4}) in two regions. In the region where the
distance between excitons is much smaller than the wavelength of their relative
motion, $kr\ll1$, the coordinate dependence of the wave function comes only
from the potential energy and the characteristic scale $r\sim d^{2}/b$. At this
scale the kinetic energy can be neglected and Eq.(\ref{eq:xxs.4}) is reduced to
\begin{equation}
\nabla^{2}\psi(\bm{r}) - \frac{d^{2}}{br^{3}} \ \psi(\bm{r}) = 0 \ ,
    \hspace{1cm} kr \ll 1 \ .
\label{eq:xxs.5}
\end{equation}
When the distance between the excitons is much larger than
$r_{0}=(e^{2}d^{2}/\kappa E)^{1/3}=(2d^{2}/k^{2}b)^{1/3}$ the interaction
energy is small compared to the kinetic energy and
\begin{equation}
\nabla^{2}\psi(\bm{r}) + k^{2}\psi(\bm{r}) = 0 \ , \hspace{1cm}
    r \gg r_{0} \ .
\label{eq:xxs.6}
\end{equation}
Due to $kr_{0}=(kd^{2}/b)^{1/3}\ll1$ the two regions overlap at $r_{0}\ll
r\ll1/k$.

Solutions to both Eqs.(\ref{eq:xxs.5}) and (\ref{eq:xxs.6}) are expressed in
Bessel functions. When $kr_{0}\ll1$ only S scattering is important and it is
enough to find angular independent solution of Eq.(\ref{eq:xxs.5}). The
solution that goes to zero at $r\rightarrow0$ is
\begin{equation}
\psi(\bm{r}) = A_{1} K_{0}\left(\frac{2d}{\sqrt{br}}\right) =
    A_{1} K_{0}\left(2kr_{0} \ \sqrt{\frac{r_{0}}{r}}\right) ,
    \hspace{1cm} kr \ll 1 \ .
\label{eq:xxs.7}
\end{equation}
The solution to Eq.(\ref{eq:xxs.6}) describing scattering is
\begin{equation}
\psi(\bm{r}) = e^{i\bm{k}_{r}\bm{r}} + A_{2} H_{0}^{(1)}(kr) \ .
\label{eq:xxs.8}
\end{equation}
Making use of asymptotes \cite{gradsteyn}
\begin{subequations}
\begin{eqnarray}
&& K_{0}(z) = - \ln\frac{z}{2} - C + O(z^{2}\ln z) \ , \hspace{1cm}
    |z| \ll 1 \ ,
\label{eq:xxs.9a} \\
&& H_{0}^{(1)}(z) =
    1 + \frac{2i}{\pi} \left(\ln\frac{z}{2} + C\right) + O(z^{2}\ln z) \ ,
    \hspace{1cm} |z| \ll 1 \ ,
\label{eq:xxs.9b}
\end{eqnarray}
\label{eq:xxs.9}
\end{subequations}
where $C=0.577$ is the Euler constant it is easy to match the solutions in the
intermediate region $r_{0}\ll r\ll1/k$:
\begin{equation}
- A_{1}\left(\ln\frac{kr_{0}^{3/2}}{r^{1/2}} + C\right) =
    1 + A_{2}
    \left[1 + \frac{2i}{\pi} \left(\ln\frac{kr}{2} + C\right)\right] .
\label{eq:xxs.10}
\end{equation}
This gives
\begin{equation}
A_{1} = - \frac{2}{\ln(kd^{2}/b) + 3C - \ln2 - i\pi/2} \ ,
    \hspace{0.5cm}
    A_{2} = \frac{i\pi/2}{\ln(kd^{2}/b) + 3C - \ln2 - i\pi/2} \ .
\label{eq:xxs.11}
\end{equation}
According to Eq.(\ref{eq:xxs.1b}) the argument of the logarithm is small and
with the logarithmic accuracy
\begin{equation}
A_{1} = - \frac{2}{\ln(kd^{2}/b)} \ ,
    \hspace{1cm}
    A_{2} = \frac{i\pi/2}{\ln(kd^{2}/b)} \ .
\label{eq:xxs.12}
\end{equation}

>From the asymptote \cite{gradsteyn}
\begin{equation}
H_{0}^{(1)}(z) = \sqrt\frac{2}{\pi z} \ e^{i(z-\pi/4)} \ , \hspace{1cm}
    |z| \gg 1 \ ,
\label{eq:xxs.13}
\end{equation}
it follows
\begin{equation}
\psi(\bm{r}) = e^{i\bm{kr}} +
    \frac{e^{i\pi/4}}{\ln(kd^{2}/b)} \sqrt\frac{\pi}{2kr} \ e^{ikr} \ ,
    \hspace{1cm} kr \gg 1 \ .
\label{eq:xxs.14}
\end{equation}
That is the scattering crosssection (in 2D case it has units of length) is
\begin{equation}
\sigma = \frac{\pi^{2}}{k\ln^{2}(kd^{2}/b)} \ .
\label{eq:xxs.15}
\end{equation}


\end{document}